# The Effect of Dielectric Crystals on the Electron Diffraction

Yulian Shabolovski

University of California, Los Angeles; Department Physics & Astronomy

yulshab95@gmail.com

A dynamic diffraction theory is developed for describing electron diffraction by dielectric crystals in a strong electromagnetic field. It is shown that additional diffraction maxima arise in an electromagnetic field, their intensity appreciably depending on the field strength. In some cases the intensity of the diffraction maxima is modulated by the electromagnetic field frequency.

We consider here the diffraction of electrons by a dielectric that is continuously illuminated by a laser. The diffraction of electrons in a crystal in the electromagnetic field of an optical laser was investigated experimentally by Schwarz and Hora [1,2]. It is not our purpose here to explain these experiments; we confine ourselves to elastic scattering of electrons (the states of the atoms of the crystal are not altered by the scattering process) in a dielectric single crystal in the field of a plane electromagnetic wave with circular polarization.

## 1. THE WAVE FUNCTION

The wave function of an electron in the field of an electromagnetic wave was calculated by Volkov see, e.g., [3])

$$\varphi_{\mathbf{q}} = \left[1 + \frac{e}{2(kp)}(\widehat{k}\widehat{a}_z \sin\varphi + \widehat{k}\widehat{a}_1 \cos\varphi\right] \frac{u(p)}{\sqrt{2q_0}} \exp\left\{-ie\frac{a_1 p}{(kp)}\sin\varphi + ie\frac{a_2 p}{(kp)}\cos\varphi - iqx\right\}, \qquad (1)$$

$$q = p - e^2 \frac{a^2}{2(kp)} k,$$

the notation is that of [2].

We construct a dynamic theory of scattering of electrons having the wave function (1) in a crystal, using a procedure similar to that used for the scattering of plane waves of electrons [4] and neutrons [5] in crystalline matter.

We represent the Hamiltonian of the interaction of the electron with the crystal in the form of a sum over individual atoms of the crystal:

$$V(\mathbf{r}) = \sum_j V_j(\mathbf{r}). \qquad (2).$$

We expand the wave function w(r, t) of the electron in the crystal in terms of the functions (1):

$$\Psi(\mathbf{r},t) = \sum_{\mathbf{q}} \Psi_{\mathbf{q}}(t)\varphi_{\mathbf{q}}(\mathbf{r},t). \qquad (3)$$

We then obtain from the Schrodinger equation $\Psi_{\mathbf{q}}(t)$ the following equation for the coefficients (h = c = 1):

$$i\dot{\Psi}_{\mathbf{q}}(t) = \sum_j \sum_{\mathbf{q}'} (V_j)_{\mathbf{qq}'} \Psi_{\mathbf{q}'}(t), \qquad (4)$$

where $(V_j)_{\mathbf{qq}'}$ in (4) is the matrix element of $V_j(\mathbf{r})$ between the functions (1) and their explicit form is given by [3]

$$(V_j)_{\mathbf{qq}'} = \sum_{s=-\infty}^{\infty} M_s(\mathbf{qq}')e^{i(\mathbf{q}-\mathbf{q}'-\mathbf{k}s)} R_{je} i(\varepsilon_{\mathbf{q}} - \varepsilon_{\mathbf{q}'} + \omega s)t, \qquad (5)$$

where

$$M_s(\mathbf{qq}') = \frac{1}{2(\varepsilon_{\mathbf{q}}\varepsilon_{\mathbf{q}'})^{1/2}} V(\mathbf{q}-\mathbf{q}'-\mathbf{k}s)[2mB_s(z) + \beta_1 B_{1s}(z) + \beta_2 B_{2s}(z)], \quad B_s(z) = J_s(z)e^{is\varphi_0},$$

$$B_{1s}(z) = \frac{1}{2}\left[J_{s+1}(z)e^{i(s+1)\varphi_0} + J_{s-1}(z)e^{i(s-1)\varphi_0}\right], \quad B_{2s}(z) = \frac{1}{2i}\left[J_{s+1}(z)e^{i(s+1)\varphi_0} + J_{s-1}(z)e^{i(s-1)\varphi_0}\right],$$

$$z^2 = \alpha_1^2 + \alpha_2^2, \quad \cos\varphi_0 = \alpha_1 / z, \quad \sin\varphi_0 = \alpha_2 / z,$$

$$\alpha_1 = e\left(\frac{aq_x}{\omega\varepsilon_{\mathbf{q}}} - \frac{aq_x'}{\omega\varepsilon_{\mathbf{q}'}}\right), \quad \alpha_2 = e\left(\frac{aq_y}{\omega\varepsilon_{\mathbf{q}}} - \frac{aq_y'}{\omega\varepsilon_{\mathbf{q}'}}\right),$$

$$\beta_1 = -\frac{e}{2m} a_x (\varepsilon_{\mathbf{q}}{}^2 - m^2)^{1/2} \left\{1 + \frac{q_x'}{|\mathbf{q}|} - i\frac{q_y'}{|\mathbf{q}|}\langle\sigma_z\rangle + \frac{\varepsilon_{\mathbf{q}} - m}{\sqrt{\varepsilon_{\mathbf{q}}{}^2 - m^2}} \frac{q_z'}{|\mathbf{q}|}\right\},$$

$$\beta_2 = \frac{e}{2m} a_y (\varepsilon_{\mathbf{q}}{}^2 - m^2)^{1/2} \left\{\left(i\langle\sigma_z\rangle + \frac{q_y'}{|\mathbf{q}|} - i\frac{q_x'}{|\mathbf{q}|}\langle\sigma_z\rangle\right) - i\frac{\varepsilon_{\mathbf{q}} - m}{\sqrt{\varepsilon_{\mathbf{q}}{}^2 - m^2}} \frac{q_z}{|\mathbf{q}|}\langle\sigma_z\rangle\right\};$$

$$\varepsilon_{\mathbf{q}} = (q^2 + m^2)^{1/2}.$$

It is assumed here that the plane electromagnetic wave is defined by a 4-potential in the form

$$A = a_1\cos\varphi + a_2\sin\varphi, \quad \varphi = -kz + \omega t.$$

We assume a Lorentz gauge for the potential, so that $((a_1k) = (a_2k) = 0$. We choose a coordinate system in which the x axis is along $a_1$, the y axis along $a_2$, and the z axis along k. We include in (5) the terms corresponding to electron-atom interactions in which the electron spin remains unchanged. If the electron spin is altered by interaction with the atom, then the resultant state is not coherent with the initial state 'and this process should be regarded as inelastic. In the absence of initial polarization of the atoms, the process characterized by interference between the incident and the scattered waves is described only by the remaining terms of (5) which do not change the electron spin. If the scattering amplitude f(qq') in the intervals of interest to us is smaller than the distance between neighboring atoms, then ([6])

$$V_j(\mathbf{q},\mathbf{q}') = -\frac{2\pi}{m} f_j(\mathbf{qq}'), \tag{6}$$

where $f_j = [f^+(I+1) + f^- I]/(2\mathrm{I}+1)$, and $f^+$ and $f^-$ are the amplitudes for the scattering of the electron by an individual atom with spin I corresponding to two values of the total spin J = I ± 1/2. To simplify the calculations that follow, we assume that the Debye-Waller factor is identically equal to unity, i.e., only scattering without electron recoil from the atoms takes place. It is easily seen that these two approximations do not affect the considered qualitative effects, but facilitate greatly the algebraic manipulations.

Summing over the index j in the second term of (4) and taking (5) into account, we obtain

$$i\dot{\Psi}_{\mathbf{q}}(t) = \sum_{\mathbf{K}}\sum_{s} M(\mathbf{q},\mathbf{q}_{\mathbf{K}_s}) \exp\left\{i(E_{\mathbf{q}_{\mathbf{K}_s}} - E_{\mathbf{q}} + \omega s)t\right\} \Psi_{\mathbf{q}_{\mathbf{K}_s}}(t), \tag{7}$$

where $K = 2\pi b$, b is the reciprocal-lattice vector, and $q_{K_s} = q + K - ks$. We seek

$$\Psi_{\mathbf{q}}(t) = U_{\mathbf{q}} e^{-i(E_{\mathbf{q}} - E)t}, \quad \Psi_{\mathbf{q}_{\mathbf{K}_s}}(t) = U_{\mathbf{q}_{\mathbf{K}_s}} \exp\left\{-i(E_{\mathbf{q}_{\mathbf{K}_s}} - E_{\mathbf{q}} + \omega s)t\right\},$$

and then the system (7) takes the form

$$\left(\frac{\mathbf{q}^2}{\chi^2} - 1\right) U_{\mathbf{q}} = \sum_{s}\sum_{\mathbf{K}} g(\mathbf{q},\mathbf{q}_{\mathbf{K}_s}) U_{\mathbf{q}_{\mathbf{K}_s}}. \tag{8}$$

Here

$$g(\mathbf{q},\mathbf{q}_{\mathbf{K}_s}) = \frac{4\pi}{\chi^2 V_0} M(\mathbf{q},\mathbf{q}_{\mathbf{K}_s})$$

$V_0$ is the volume of the unit cell and $k = (2mE)^{1/2}$. It is assumed in (8) that the crystal consists of identical atoms. If this is not the case, then expression (6) for Vj should be replaced by the isotropicaly coherent part of the scattering amplitude.

We examine the system (8) and analyze first the simplest case, when the wave vector ql has for one of the diffracted waves is close to the value obtained from the exact Bragg condition, $q_1 = q + K - ks$ . We can then separate from (8) only two equations, which are the same for both values of the electron polarization:

$$\left(\mathbf{q}^2/\chi^2 - 1\right)U_{\mathbf{q}} = g_{00}U_{\mathbf{q}} + g_{01}U_{\mathbf{q}_1},$$
$$\left(\mathbf{q}_1{}^2/\chi^2 - 1\right)U_{\mathbf{q}_1} = g_{01}U_{\mathbf{q}} + g_{11}U_{\mathbf{q}_1}. \tag{9}$$

The system (9) is analogous to the system considered in [5], and we therefore seek its solution in the same form.

We consider a crystal in the form of a flat plate in the Laue case (the diffracted beam passes through the crystal). We then have for the wave function of an electron with arbitrary polarization in the crystal [5]

$$\Psi(\mathbf{r},t) = \Phi_0 e^{i\chi\mathbf{r}} e^{-iE_\chi t} - \left[\left(\frac{2\varepsilon_0^{(2)} - g_{00}}{2(\varepsilon_0^{(2)} - \varepsilon_0^{(1)})}\mathrm{E}^{(1)} - \frac{2\varepsilon_0^{(1)} - g_{00}}{2(\varepsilon_0^{(2)} - \varepsilon_0^{(1)})}\mathrm{E}^{(2)}\right)\varphi_\chi\right.$$
$$\left. + \frac{g_{10}\beta}{2(\varepsilon_0^{(2)} - \varepsilon_0^{(1)})}(\mathrm{E}^{(1)} - \mathrm{E}^{(2)})e^{i\mathbf{K}\mathbf{r} - i\mathbf{k}_s\mathbf{r} + i\omega st}\varphi_{\chi_i}\right],$$
$$\varepsilon_0^{(1)(2)} = 1/4(g_{00} + g_{11}\beta - \alpha\beta - 2ms\omega\beta/\chi^2) \pm \tag{10}$$
$$\pm 1/4[(g_{00} + g_{11}\beta - \alpha\beta - 2ms\omega\beta/\chi^2)^2 + 4\beta(\alpha + 2ms\omega/\chi^2) - \Delta]^{1/2},$$
$$\mathrm{E}^{(1)(2)} = \exp\left\{i\varepsilon^{(1)(2)}\mathbf{n}\mathbf{r}/\gamma\right\}, \quad \Delta = g_{00}g_{11} - g_{01} + g_{10};$$

$\alpha = (K - ks)(K - ks + 2k)/k^2$ is the angle characterizing the deviation from the Bragg conditions; $\beta = \gamma_0/\gamma_1$; $\gamma_{0,1} = \cos\theta_{0,1} = \mathbf{n}\mathbf{k}_{0,1}$ , $\mathbf{k}_1 = \mathbf{k} + \mathbf{K} - \mathbf{k}s$; n is the inward normal to the face; $\Phi_0$ is the value of the wave function of the electron at the entrance surface of the crystal.

A detailed analysis of the solution will be presented below; we note here only the qualitative difference between the diffraction of the electron in the electromagnetic wave and ordinary diffraction. Comparing the Bragg condition without the field, $\mathbf{q}_1 = \mathbf{q}_0 + \mathbf{K}$ q, with the analogous condition in the field, $\mathbf{q}_1 = \mathbf{q}_2 + \mathbf{K} - \mathbf{k}s$ , we see that in addition to the usual direction of the diffracted wave at s = 0 there are produced in the laser field many other waves in directions determined by the Bragg condition $\mathbf{q}_1 = \mathbf{q}_0 + \mathbf{K} \pm s\mathbf{k}$ q (s = 1, 2, 3, ... ). The intensity of these new scattered waves depends strongly, of course, on the electromagnetic field strength, and will be estimated later on. In addition, in scattering by a crystal it is possible to have a case in which several diffracted waves propagate in one direction. For example, we consider the case when the waves $\mathbf{q}_1 = \mathbf{q}_0 + \mathbf{K} + s_1\mathbf{k}$ and $\mathbf{q}_1 = \mathbf{q}_0 + \mathbf{K}_2 - s_2\mathbf{k}$ propagate in the direction of $\mathbf{q}_1 = \mathbf{q}_0 + \mathbf{K}$; this is obviously possible of the following condition is satisfied:

$$\mathbf{K}_0 = \mathbf{K}_1 + s_1\mathbf{k} = \mathbf{K}_2 - s_2\mathbf{k} \tag{11}$$

where $\mathbf{K}_0$, $\mathbf{K}_1$, and $\mathbf{K}_2$ are different reciprocal-lattice vectors multiplied by $2\pi$.

## 2. THE INTENSITY OF THE SCATTERED WAVE

Let us examine this case in greater detail, and assume for simplicity that a second wave with $\mathbf{q}_2 = \mathbf{q}_0 + \mathbf{K}_1 + s\mathbf{k}$ propagates in the direction of one diffracted wave $\mathbf{q}_1 = \mathbf{q}_0 + \mathbf{K}$, with $\mathbf{q}_1 = \mathbf{q}_2$. The system (8) then takes the form

$$\begin{aligned}
&\left(\mathbf{q}_0{}^2/\chi^2 - 1\right)U_{\mathbf{q}_0} = g_{00}U_{\mathbf{q}_0} + g_{01}U_{\mathbf{q}_1} + g_{02}U_{\mathbf{q}_2},\\
&\left(\mathbf{q}_1{}^2/\chi^2 - 1\right)U_{\mathbf{q}_1} = g_{10}U_{\mathbf{q}_0} + g_{11}U_{\mathbf{q}_1} + g_{12}U_{\mathbf{q}_2}.\\
&\left(\mathbf{q}_2{}^2/\chi^2 - \frac{2m\omega}{\chi^2}s - 1\right)U_{\mathbf{q}_2} = g_{20}U_{\mathbf{q}_0} + g_{21}U_{\mathbf{q}_1} + g_{22}U_{\mathbf{q}_2}
\end{aligned} \tag{12}$$

Let $\mathbf{q} = \mathbf{k} + k_\delta\mathbf{n}$, $q_0 = (1+\varepsilon_0)\,\mathrm{k}$, $q_1 = (1+\varepsilon_1)k$, and $q_2 = (1+\varepsilon_2)k$; we have then in analogy with [5] $\delta = \varepsilon_0/\gamma_0$, where $\gamma_0 = \cos(\mathbf{q}_0 \cdot \mathbf{n})$, and from the condition $\mathbf{q}_1 = \mathbf{q}_0 + \mathbf{K}$ we obtain

$$\varepsilon_1 = \frac{\alpha}{2} + \frac{\varepsilon_0\gamma_1}{\gamma_0}, \quad \alpha = \frac{\mathbf{K}_0(\mathbf{k}_0 + 2\boldsymbol{\chi}}{\chi^2};$$

from $\mathbf{q}_2 = \mathbf{q}_0 + \mathbf{K}_1 + s\mathbf{k}$ we obtain analogously

$$\varepsilon_2 = \frac{\alpha_1}{2} + \frac{\varepsilon_0\gamma_2}{\gamma_1}, \quad \alpha_1 = \frac{(\mathbf{K}_1 + s\mathbf{k})(\mathbf{K}_1 + s\mathbf{k} + 2\boldsymbol{\chi})}{\chi^2},$$

where, by definition, $\gamma_2 = \cos(\mathbf{q}_2 \cdot \mathbf{n}) \equiv \gamma_1$. Substituting the obtained relations in (12) and confining ourselves to the first order in E, we have

$$\begin{aligned}
&\left(2\varepsilon_0 - g_{00}\right)U_{\mathbf{q}_0} - g_{01}U_{\mathbf{q}_1} - g_{02}U_{\mathbf{q}_2} = 0,\\
&\left(2\varepsilon_0\gamma_1/\gamma_0 + \alpha - g_{11}\right)U_{\mathbf{q}_1} - g_{10}U_{\mathbf{q}_0} - g_{12}U_{\mathbf{q}_2} = 0,\\
&\left(2\varepsilon_0\gamma_1/\gamma_0 + \alpha_1 - \frac{2m\omega}{\chi^2}s - g_{22}\right)U_{\mathbf{q}_2} - g_{20}U_{\mathbf{q}_0} - g_{21}U_{\mathbf{q}_1} = 0.
\end{aligned} \tag{13}$$

From the condition of the compatibility of the system (13) we obtain an equation for E:

$$\begin{aligned}
&\left(2\varepsilon_0 - g_{00}\right)(2\varepsilon_0 + \beta\alpha - \beta g_{11})(2\varepsilon_0 + \beta\alpha_1 - \beta 2m\omega s/\chi^2 - \beta g_{22}) \\
&-\beta^2 g_{21}g_{10}g_{02} - \beta^2 g_{01}g_{12}g_{20} - \beta g_{20}g_{02}(2\varepsilon_0 + \beta\alpha - \beta g_{11}) - \\
&-\beta^2 g_{21}g_{12}\left(2\varepsilon_0 - g_{00}\right) - \beta g_{10}g_{01}(2\varepsilon_0 + \beta\alpha_1 - \beta 2m\omega s/\chi^2 - \beta g_{22}) = 0
\end{aligned} \tag{14}$$

Equations (13) and (14), with the following boundary conditions on the entrance surface of the crystal

$$\begin{aligned}
&U_{\mathbf{q}_0}{}^{(1)}(0) + U_{\mathbf{q}_0}{}^{(2)}(0) + U_{\mathbf{q}_0}{}^{(3)}(0) = \Phi_0, \\
&U_{\mathbf{q}_1}{}^{(1)}(0) + U_{\mathbf{q}_1}{}^{(2)}(0) + U_{\mathbf{q}_1}{}^{(3)}(0) = 0, \\
&U_{\mathbf{q}_2}{}^{(1)}(0) + U_{\mathbf{q}_2}{}^{(2)}(0) + U_{\mathbf{q}_2}{}^{(3)}(0) = 0.
\end{aligned} \tag{15}$$

($\Phi_0$ is the function incident on the crystal boundary), determine completely the wave function of the electron in the crystal

$$\begin{aligned}
&\Psi(\mathbf{r},t) = \Phi_0[\Psi_0(z)\varphi(\chi_0) + \Psi_1(z)\varphi(\chi_1) + \Psi_2(z)\varphi(\chi_2)\,\mathrm{e}^{-is\omega t}]e^{iE_\chi t}, \\
&\Psi_0(z) = \Delta^{-1}(A\mathrm{E}_1(z) + B\mathrm{E}_2(z) + C\mathrm{E}_3(z)), \\
&\Psi_1(z) = \Delta^{-1}(a_1 A\mathrm{E}_1(z) + a_2 B\mathrm{E}_2(z) + a_3 C\mathrm{E}_3(z)), \\
&\Psi_2(z) = \Delta^{-1}(b_1 A\mathrm{E}_1(z) + b_2 B\mathrm{E}_2(z) + b_3 C\mathrm{E}_3(z)), \\
&\Delta = (a_2 b_3 - a_3 b_2) - (a_1 b_3 - a_3 b_1) + (a_1 b_2 - a_2 b_1) = A + B + C, \\
&a_i = \left(1 - \frac{2\varepsilon_0^{(i)} - g_{00}}{g_{01}}\right) \Big/ \left(1 + \frac{g_{02}}{\beta g_{12}}\frac{2\varepsilon_0^{(i)} + \beta\alpha - \beta g_{11}}{g_{01}}\right), \\
&b_i = \left(\frac{g_{01}}{g_{02}} + \frac{2\varepsilon_0^{(i)} - g_{00}}{g_{01}}\frac{2\varepsilon_0^{(i)} + \beta\alpha - \beta g_{11}}{g_{01}}\right) \Big/ \left(1 + \frac{g_{02}}{\beta g_{12}}\frac{2\varepsilon_0^{(i)} + \beta\alpha - \beta g_{11}}{g_{01}}\right), \\
&z = \mathbf{nr}, \quad E_1(z) = \exp\left\{-i\chi\varepsilon_0^{(i)}/\gamma_0\right\}, \quad l = 1,2,3,
\end{aligned} \tag{16}$$

and $\varepsilon_0^l$ are the roots of (14).

As seen from (16), the intensity of the scattered wave is in this case

$$I_{pac} = \left|\Phi_0\right|^2 \left|\Psi_1(z)\varphi(\chi_1) + \Psi_2(z)\varphi(\chi_2)\,\mathrm{e}^{-is\omega t}\right|^2. \tag{17}$$

We note immediately one feature of formula (17) – the intensity of the diffracted electrons is modulated both in terms of time and in terms of coordinate in the direction of wave propagation.

The exact expression (17) for the intensity of the scattered electrons is quite cumbersome. We therefore confine ourselves for simplicity to the diffraction of
electrons in a weak electromagnetic field, when absorption and emission of one photon by the electron are the most probable. We consider the appearance of additional Bragg maxima near some diffraction maximum that exists also without an electromagnetic field. We choose, for example, the zeroth Bragg maximum. Then two additional maxima can be produced near it in a weak electromagnetic field, and correspond to absorption (s = -1) and emission (s = +1) of one photon. The amplitudes of the scattered waves in these maxima are given by (10) if the maxima are well resolved, i.e., the angle distance ~ k/ q between them is much large r than the dimension $\sim (qD)^{-1}$ of the diffraction maximum. Here D is the diameter of the electron beam if it is smaller than the transverse dimension of the crystal; otherwise it is necessary to substitute for D the characteristic dimension of the crystal. The intensity at the new maxima is equal to

$$I = \left|\frac{g_{01}\beta}{2(\varepsilon_0^{(2)} - \varepsilon_0^{(1)})}\right|^2 \left|e^{i\varepsilon_0^{(1)} z/\gamma_0} - e^{i\varepsilon_0^{(2)} z/\gamma_0}\right|^2, \qquad (18)$$

where $\varepsilon_0^{(1)} \approx (1/2)(g_{00} + g_{11}\beta - 2m\omega\beta / k^2)$, $\varepsilon_0^{(2)}$ ~ 0, and $\Delta$= o if $\alpha \approx 0$, i.e., if the Bragg condition is satisfied exactly. If the crystal is also thin enough, $\varepsilon_0^{(1)(2)} L/\gamma_0 << 1$ (L is the crystal thickness), then

$$I \cong \left|g_{01}\beta L / 2\gamma_0\right|^2, \qquad (19)$$

where $g_{01} \approx g_{01}{}^{(0)} J_1(\alpha)\exp(i\varphi_0)$, $\alpha^2 = 2\xi^2 \frac{m}{\omega}$, $\xi^2 = \frac{e^2 a^2}{m^2}$,

$$g_{01}{}^{(0)} = \frac{4\pi}{\chi^2 V_0} f(\mathbf{q}_0\mathbf{q}_1), \quad g_{00} \approx g_{00}{}^{(0)} J_0(\alpha) e^{i\varphi_0}, \quad g_{11} \approx g_{11}{}^{(0)} J_0(\alpha) e^{i\varphi_0},$$
$$g_{00}{}^{(0)} = \frac{4\pi}{\chi^2 V_0} f(\mathbf{q}_0\mathbf{q}_0), \quad g_{11}{}^{(0)} = \frac{4\pi}{\chi^2 V_0} f(\mathbf{q}_1\mathbf{q}_1).$$

In the other limiting case, when all three maxima overlap, the amplitudes of the three modes must be added to obtain the intensity of the diffracted beam. We then have for a thin crystal

$$I = \left|g_{00} e^{i\mathbf{q}_0\mathbf{r}} + \beta g_{10} e^{i(\mathbf{q}_0+\mathbf{k})\mathbf{r} - i\omega t} - g_{10}{}'\beta e^{i(\mathbf{q}_0-\mathbf{k})\mathbf{r} + i\omega t}\right|^2 \left(\frac{\chi L}{2\gamma_0}\right)^2, \qquad (20)$$

or, assuming that g $g_{10}{}^0 \sim g_{00}{}^0 g_{11}{}^0$ we obtain from (20)

$$I \approx \left|g_{00}\right|^2 (1 + \beta^2 (J_1^2 + J_{-1}^2) + 2\beta (J_1 + J_{-1}) \cos\theta + \beta^2 J_1 J_{-1} \cos 2\theta,$$
$$\theta = kz - \omega t + \varphi_0. \qquad (21)$$

Thus, as seen from (21), oscillations both with respect to time and with respect the distance from the crystal appear in the expression for the intensity of the scattered electrons in the case when three diffracted waves overlap.

## 3. CONCLUSION

The picture of the diffraction of electrons by a crystal in a strong electromagnetic field is thus as follows: First, additional Bragg maxima are produced in the direction $\mathbf{q}_s = \mathbf{q}_0 + \mathbf{K} \pm s\mathbf{k}$; these maxima do not exist in the absence of the electromagnetic field. They are well resolved; for example, the angle distance between them is ~k/q, which is much larger than the angular dimension $\sim (qD)^{-1}$ of the diffraction maximum (D was defined earlier). Second, several waves can now propagate without loss in the Bragg directions for the scattered wave without the field; this leads to time modulation of the electron intensity in these directions, with frequency *sw,* and the modulation depends on the field and becomes noticeable in fields on the order of $\xi$ ~1(see also [8-75]).

We did not touch upon the question of detection of the diffraction electrons. This problem is quite complicated and requires a detailed analysis, which will be published separately.